\documentclass[final, 5p, two column,times, sort&compress]{elsarticle}
\usepackage{graphicx}
\usepackage{subfigure}
\usepackage{gensymb}
\usepackage{booktabs}
\usepackage{amssymb}
\usepackage{natbib}

\biboptions{square,super,comma,numbers}

\begin{document}

\begin{frontmatter}

\title{On the Application of Viscoelastic \& Viscoplastic Constitutive Relations\\  in the CFD Bio-Fluid Simulations}

\author{S. M. Javid Mahmoudzadeh Akherat}

\address{Mechanical, Materials and Aerospace Engineering Department, Illinois Institute of Technology, Chicago, Illinois, USA}

\begin{abstract}

Considerations on implementation of the stress-strain constitutive relations applied in Computational Fluid dynamics (CFD) simulation of cardiovascular flows have been addressed extensively in the literature. However, the matter is yet controversial. The author suggests that the choice of non-Newtonian models and the consideration of non-Newtonian assumption versus the Newtonian assumption is very application oriented and cannot be solely dependent on the vessel size.  In the presented work, where a renal disease patient-specific geometry is used, the non-Newtonian effects manifest insignificant, while the vessel is considered  to be medium to small which, according to the literature, suggest a strict use of non-Newtonian formulation. The insignificance of the non-Newtonian effects specially manifests in Wall Shear Stress (WSS) along the walls of the numerical domain, where the differences between Newtonian calculated WSS and non-Newtonian calculated WSS is barely visible.

\end{abstract}


\end{frontmatter}

\section{Introduction}

In this article, the application of two non-Newtonian constitutive relations applied to a complex two-dimensional geometry is addressed, as that is one f the first decisions to make while tackling a CFD simulation of patient-specific geometries in biomedical practice. 

 In general, blood is a multiphase combination of liquid plasma as the solvent medium and platelets, deformable red blood cells, white blood cells, and other solid particles in the solution. Although this is very well known, the choice of non-Newtonian assumption has been a center of controversy in the field of hemodynamics for decades and many authors have contradictory approaches toward this issue. It is very well known that blood is a shear-thinning fluid. At higher shear rates, blood acts very similar to a Newtonian fluid while for the lower shear rates, it exhibits non-Newtonian characteristics with drastic alterations in viscosity.

Nevertheless, there are many others who believe that the non-Newtonian nature of blood has to be considered in any geometry under any condition. Taylor and Humphrey \citep{taylor} mentioned the potential importance of non-Newtonian effects on blood flow in intricate geometries, for instance,aneurysmal flows and flows in medical devices. Abraham and Behr \citep{abraham}, Barnes~\citep{barnes}, Perktold et. al. \citep{perktold3}, Leuprecht and Perktold \citep{leuprecht}, Perktold, Resch and Florian \citep{perktold1} are for the Newtonian flows, while Liepsch and Moravec \citep{moravec}, Cho and Kensey \citep{kensey},  van Wyk et. al. \citep{plesniak}, Johnston et. al. \citep{johnston}, and others have reported differences in the two assumptions.

Given the specific attention, nowadays, on the high fidelity numerical simulation exploitation in medical practice, it is beneficial to revisit the non-Newtonian flows with recent advances in CFD solvers and post-processors \cite{me1,me2,me3,me4,me5}. 

\section{Rheological Analysis}

Shear rate, $\dot{\gamma}$, is the most important hemorheological parameter in the blood flow simulation context. This is due to the fact that blood is pseudoplastic or shear-thinning, that is, the apparent viscosity decreases as the shear rate increases \citep{ws}.

Shear rate is the rate at which blood layers move along each other, measured in reciprocal seconds (1/sec)
\begin{equation}
 \dot{\gamma}=2\sqrt{D_{II}},
\label{gammadot}   
\end{equation}
in which $D_{II}$ is the second invariant of the strain rate tensor. In a two dimensional domain after simplification one will have:
\begin{equation}
D_{II}=\sum_{i,j=1}^{2}{S_{ij}S_{ij}},
\label{second invariant}   
\end{equation}
where $S_{i,j}$ is the symmetric part of the velocity gradient tensor. In tensor notation:
\begin{equation}
S_{ij}=\frac{1}{2}\left( \frac{\partial u_i}{\partial x_j}+\frac{\partial u_j}{\partial x_i} \right)   \qquad i,j=1,2.
\label{sij}   
\end{equation}

The shear rate varies from 0 to more than $1000~s^{-1}$ during every cardiac cycle and is closely related to the velocity gradient. In addition to shear rate, hematocrit is another important parameter in this context. Hematocrit (denoted here by $Hct$) is the volume percentage of the RBC in blood, which plays an integral part in determining its physiological behavior. The blood, in fact, is an emulsion of RBC in a solvent medium which is known as plasma. Plasma is mostly made of water. Hence, one can imagine that plasma acts very similar to a Newtonian fluid. There is a direct relation between Hct increment and the increase in  blood's apparent viscosity \citep{ws}\citep{quemada2}. This increase is due to roulaux density, cell-cell interaction etc. \citep{caro}. Viscometry data of a blood sample used for the our previous study (\cite{thesis}) is used for this analysis. In order to capture the non-Newtonian effects of blood, these viscosity data are then fitted to appropriate non-Newtonian models with the aim of determining the Hemorheological Parameters (HRP). These parameter were obtained using an evolutionary algorithm available in MATHEMATICA V.10. These values presented in the ensuing section.

\subsection{Constitutive Models}

Two non-Newtonian constitutive models have been considered, including Quemada and Casson.

\begin{table}[h!]
  \begin{center}
    \caption{Viscosity models considered for patient-specific hemodynamics. In this notation, $\phi=\mbox{Hct}$. Consult \citep{thesis}.}
    \label{models}
    \begin{tabular}{cc}
      \toprule
      Constitutive model   & Effective $\mu$\\
      \midrule
      
        \vspace{2 mm}
      Newtonian fluid & constant patient-specific viscosity, $\mu$\\
      
          \vspace{2 mm}
   Quemada & $\mu\left(\dot{\gamma},\phi\right)=\mu_{F}\left(1-\frac{1}{2}\frac{k_{0}+k_{\infty}~\dot{\gamma}_{r}^{\frac{1}{2}}}{1+\dot{\gamma}_{r}^{\frac{1}{2}}}     ~\phi\right)^{-2}$  \vspace{2 mm} \\

      \vspace{2 mm}
      Casson & $\mu\left(\dot{\gamma},\tau_{y}\left(\phi\right)\right)=\left[\left(\eta^{2}\frac{\mid\dot{\gamma}\mid^{2}}{4}\right)^{1/4}+\sqrt{\frac{\tau_{y}\left(\phi\right)}{2}}\right]^{2}\frac{2}{\left|{\dot{\gamma}}\right|}$, \vspace{-2.5 mm} \\ &  \scriptsize{where, $\tau_{y}=0.1\left(0.625\mbox{Hct}\right)^3$ and $\eta=\mu_{F}\left(1-\mbox{Hct}\right)^{-2.5}$.} \vspace{3 mm} \\

      \bottomrule
    \end{tabular}
  \end{center}
\end{table}
In table~\ref{models}, $\mu_{F}$ is the solvent medium (plasma) viscosity which is a patient-specific quantity but varies very little from one patient to another. The average value for $\mu_{F}$ used in this investigation is $1.3\times 10^{-3}$ Pa.S.  $\dot{\gamma}_{r}$ is the reduced shear rate equal to $\frac{\dot{\gamma}}{\dot{\gamma}_{c}}$, in which ${\dot{\gamma}_{c}}$ is called the critical shear rate defined by a phenomenological kinetic model \citep{quemada2,kleinstreuer2}. Here, however, $\dot{\gamma}_{c}$ is taken to be equal to $\frac{U}{l}$, which is the ratio between the inlet velocity and the characteristic length \footnote{Similar to Neofytou's assumption in \citep{neofytou}.}. $k_{0}$ and $k_{\infty}$ are the lower and upper Quemada limit viscosities, respectively, which are related to the solute particle shape (RBC in this context) and to the structural state of the system. Casson model can be tailored for patient-specific applications only by substituting desired values for hematocrit and plasma viscosity.

\begin{figure}
  \centering{\includegraphics[scale=0.5]{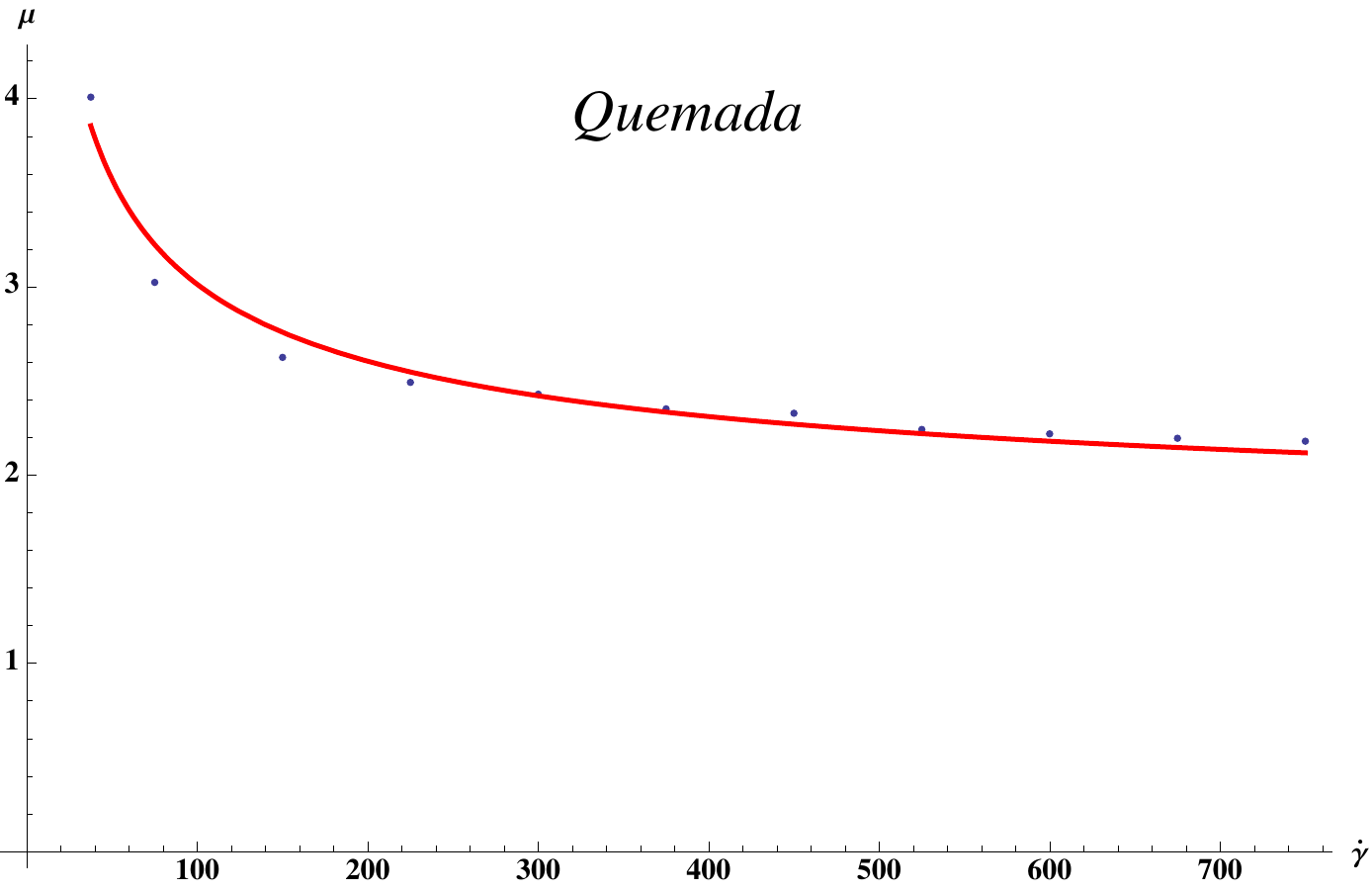}}   
  \caption{Curve fit to the viscometry data with Quemada model. shear rate and viscosity are non-dimensional here.}        
  \label{ch2 quemada35}                  
\end{figure}

 \begin{table}[h]
 \caption{Quemada parameters.}  
 \centering                          
 \begin{tabular}{c c}          
 \hline\hline                        
 Parameter & Value\\ [0.5ex]
 \hline                                      
 $k_{0}$  & 3.89\\      
 $k_{\infty}$ & 1.06\\
 $\phi$ & 0.391\\
 $\dot{\gamma}_{c}$ ($\mbox{Sec}^{-1}$) & 40.75\\[1ex] 
 \hline                                       
 \end{tabular}
 \label{ch2 quemada parameters}
 \end{table}

\section{Spectral Elements Computational Fluid Dynamics}

The CFD solver implemented here is NEK5000 spectral elements code. A patient-specific geometry of a renal failure patient vein is used here as shown in \ref{subject27geometry}. The spectral-elements are further divided into sub-elements based on Gauss-Lobato points. Producing the mesh, a polynomial order is chosen by the user based on the accuracy required to solve the problem. In this case, a polynomial order of nine was chosen. Therefore, the degrees of freedom achieved is the product of the number of raw elements and the polynomial order squared. In the case considered here, the initial element number is chosen to be 5000 with the polynomial order of seven resulting in roughly 250,000 degrees of freedom to be solved. Grid independence was verified for the numerical results. 
Conservation of mass and momentum are solved numerically and energy equation is neglected. As for the post-processing, VISIT 2.7.3 (developed by the Lawrence Livermore National Laboratory, Livermore, CA), which is an interactive parallel visualization and graphical analysis tool, was used to visualize the CFD results. The simulation starts from rest. Blood flow in the cephalic vein is laminar and can taken to be almost steady under physiological condition. The inlet velocity for this case was taken to be 0.84 [m/s].

  \begin{figure}
  \centering{\includegraphics[scale=0.35]{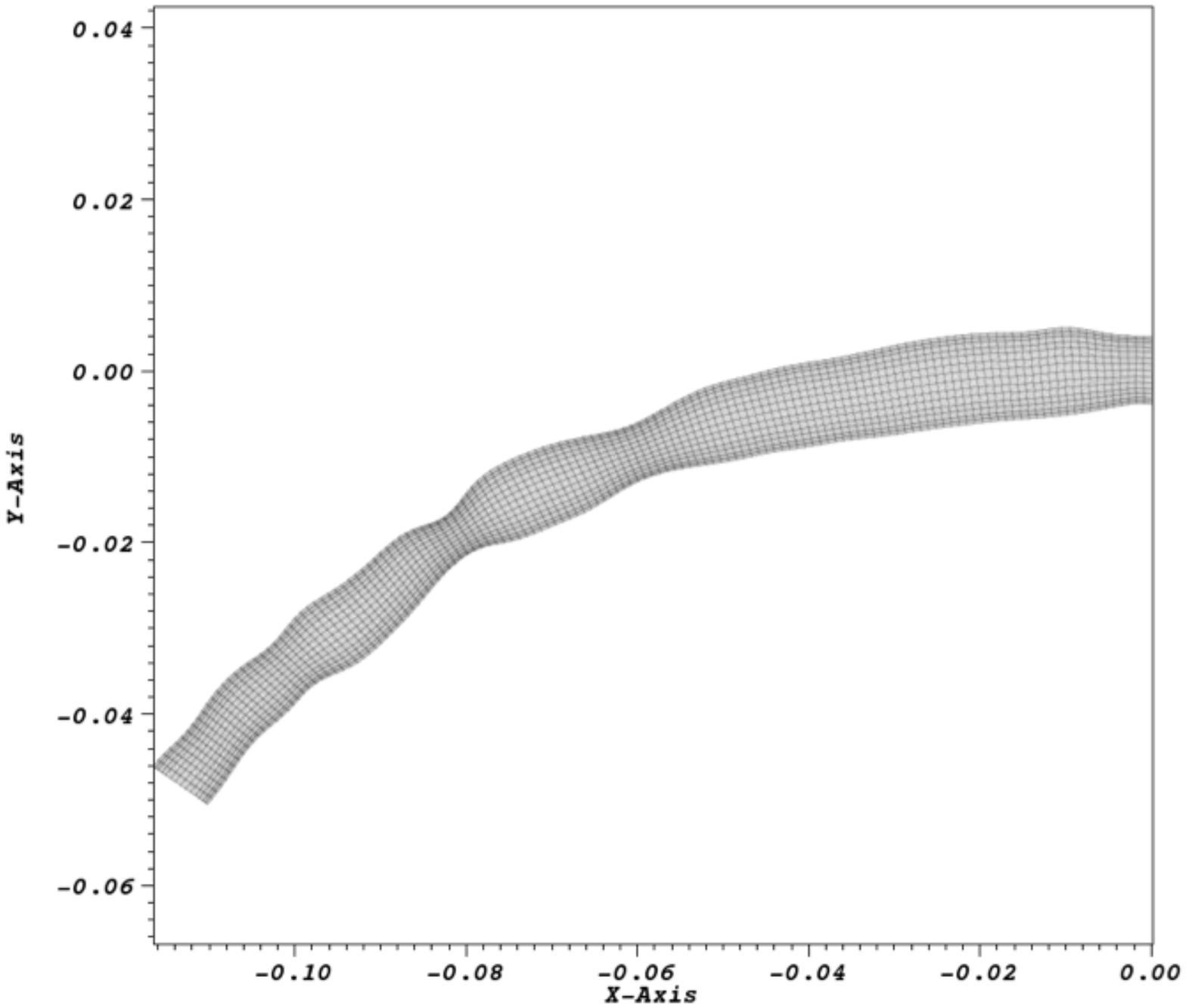}}   
  \caption{Geometry used for the CFD analysis.}       
  \label{subject27geometry}                
\end{figure}

\begin{equation}
\rho\left(\partial_{t} \textbf{u} + \textbf{u}\cdot \nabla \textbf{u}\right)= -\nabla p + \nabla \cdot \left[\mu\left(\nabla \textbf{u}+\left( \nabla \textbf{u}\right )^{T} \right) \right] + \rho \textbf{f},
\label{momentum}   
\end{equation}
\begin{equation}
\nabla \cdot \textbf{u} = 0,
\label{continuity}   
\end{equation}
in which, $\rho$ is density, $\textbf{u}$ is the fluid velocity vector, $p$ is the pressure, $\mu$ is the apparent viscosity which is the focus of this study and $\textbf{f}$ is the summation of body forces acting on the fluid. The third term on the right-hand side is the contribution to viscous effects that arises from non-Newtonian effects. For Newtonian cases, this term is set to zero in the non-stress formulation. For the non-Newtonian cases, the constitutive equation for Quemada, Walburn-Schneck, or Casson is used for the viscosity in the stress formulation.

Walls are taken to be impermeable and rigid with no-slip boundary condition applied throughout the domain. A zero pressure Dirichlet boundary condition is applied at the outlet.

\subsection{CFD Results and Discussions}

The differences between the aforementioned Newtonian and non-Newtonian assumptions can be perceived through analysis of the CFD results. Results show insignificant differences between the three categories of simulation. As for the quantitative difference between Newtonian and Quemada, the values in percent format are given in table~\ref{Newtonian and non-Newtonian Differences}. The negative sign indicates that the value corresponding to the considered parameter has increased with the implementation of the non-Newtonian model. Moreover, The variation of non-Newtonian Quemada model is illustrated through the domain in figure

  \begin{figure}
  \centering{\includegraphics[scale=0.3]{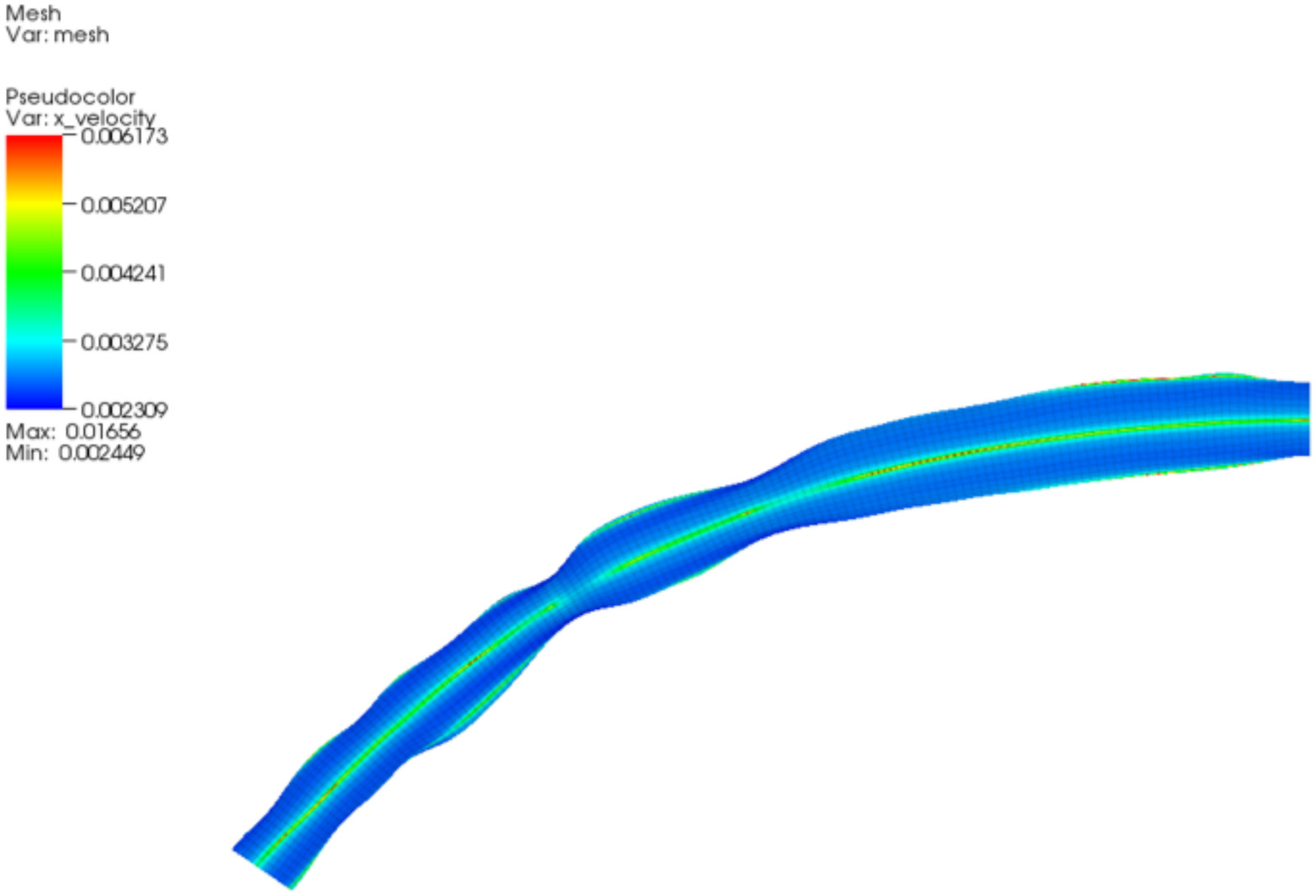}}   
  \caption{Geometry used for the CFD analysis.}       
  \label{subject27geometry}                
\end{figure}

\begin{table}[!t]
\centering
\begin{tabular}{|l||c|} \hline
{Physical Parameter} & {Difference Value} \\ \hline \hline
Max Pressure Difference   & -29\%\\  \hline
Min Pressure Difference  & 42.7\%\\  \hline
Max Vorticity Difference & -17.9\%\\  \hline
Min Vorticity Difference  & -21.3\%\\  \hline
Max X Velocity Difference & 27.5\%\\  \hline
Min X Velocity Difference & 1.5\%\\  \hline
Max Velocity Magnitude Difference & 1.4\%\\  \hline
\end{tabular}
\caption{Differences in fluid mechanics parameters between CFD results for categories two (Newtonian) and six (patients specific Quemada).}
\label{Newtonian and non-Newtonian Differences}
\end{table}

The results are comparable qualitatively except for the two recirculation zones that appear on the upper wall in the Newtonian flow that do not exist in the corresponding non-Newtonian flow. That is, a non-Newtonian flow is less prone to form recirculation zones near walls and less eddies in general. Following the same line of reasoning, one would expect that owing to the overall increase in viscosity, the reversed flows occurring in a non-Newtonian flow field would be smaller in size and slower in velocity than that of a Newtonian flow. This is clearly detectable in recirculation zones occurring in both simulations. Although a separation occurs at roughly the same spot in both simulations, the recirculation zones in the non-Newtonian simulation are smaller in length. This is in good agreement with the results of Forrester and Young \citep{forrester}. They also reported that the length of the separated flow region of blood was smaller than the one for water in their experiments.

\begin{figure}
\centering

 \subfigure[]{\includegraphics[scale=0.3]{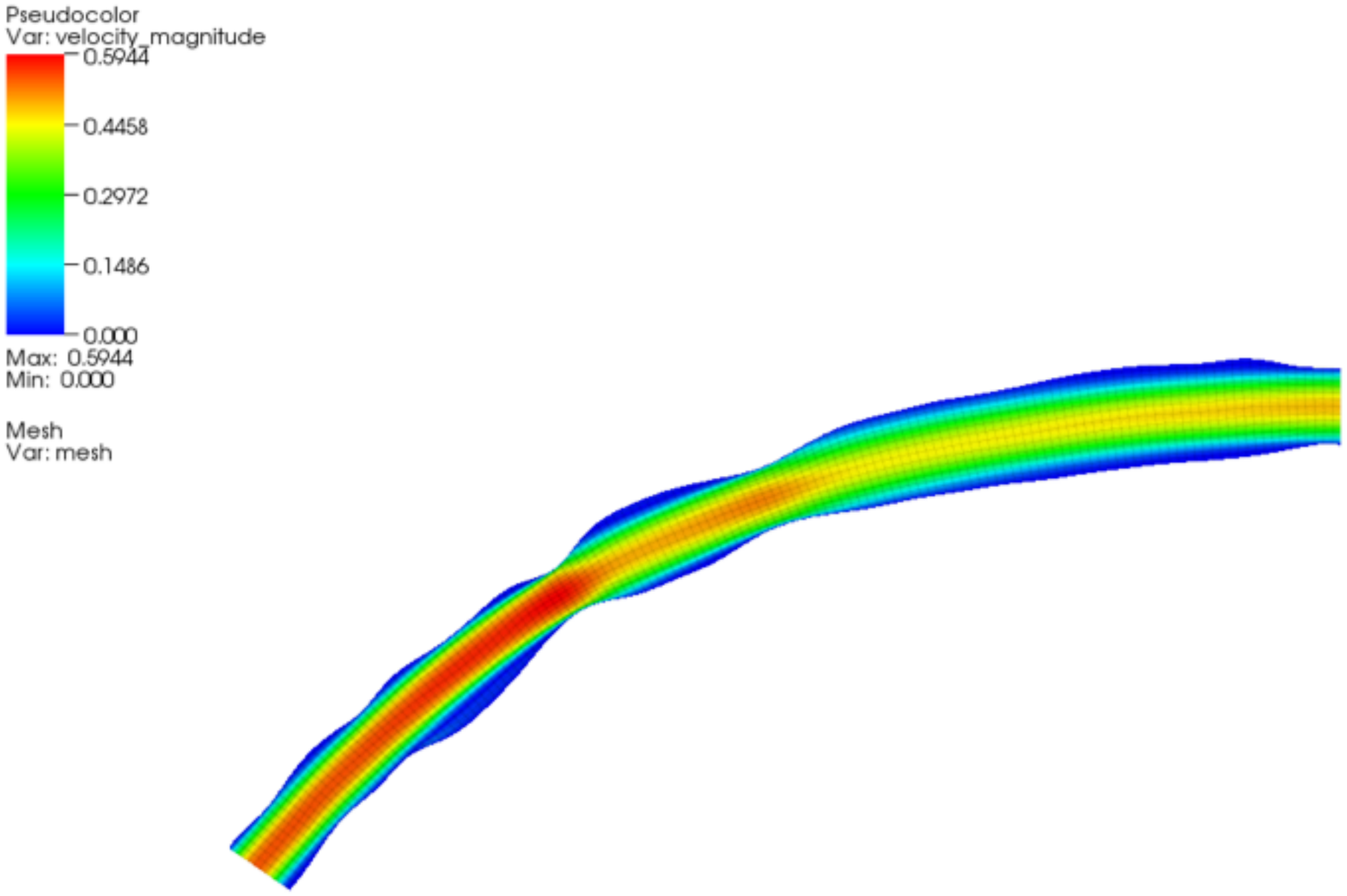}}
 \subfigure[]{\includegraphics[scale=0.3]{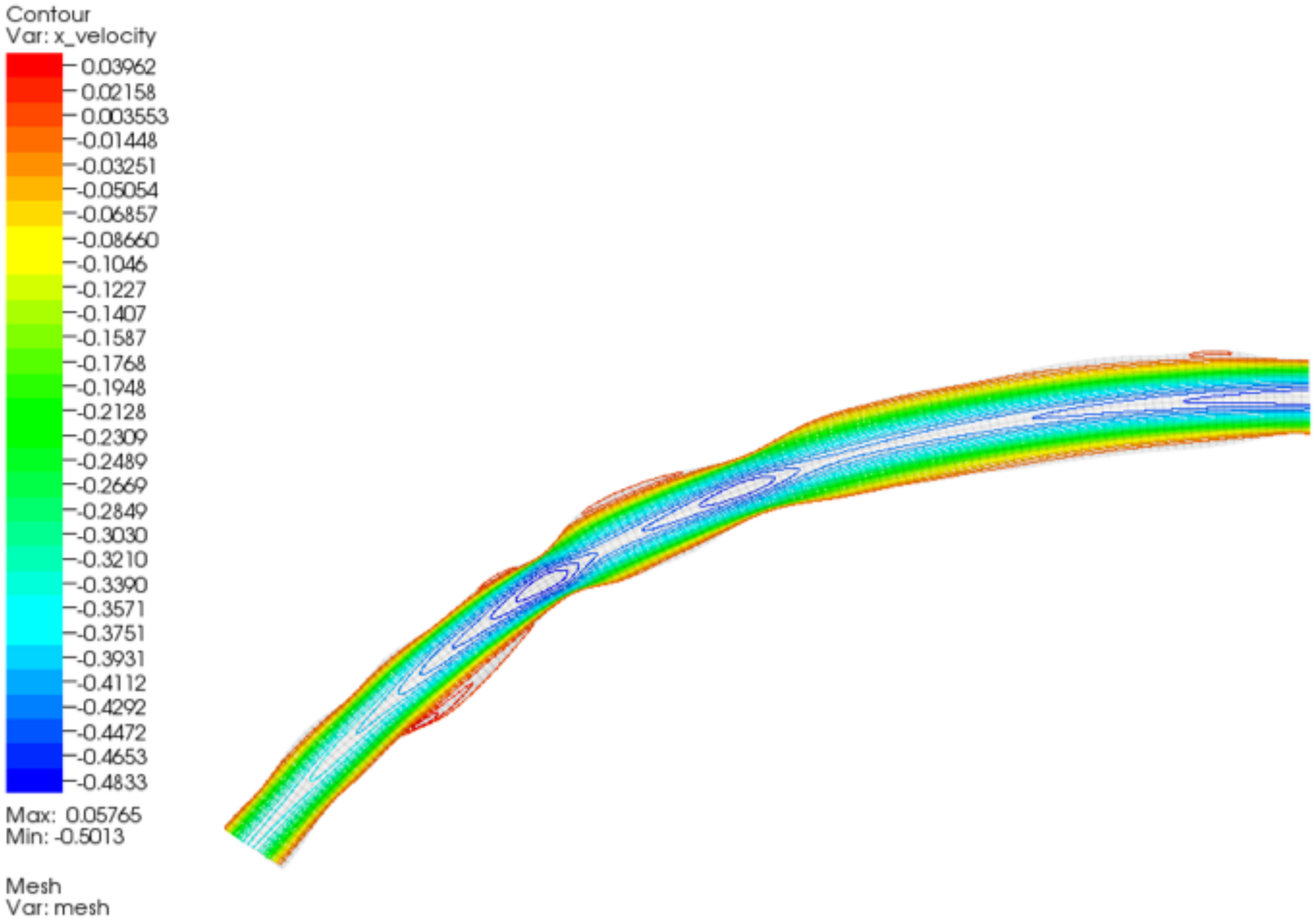}}
 \caption{(a) X velocity and (b) velocity magnitude contours - Newtonian simulation with $\mu=2.69$ cP at $t=1~\mbox{sec} $~for subject 27 at three month.}
 \label{category2}
\end{figure}

\begin{figure}
\centering

 \subfigure[]{\includegraphics[scale=0.3]{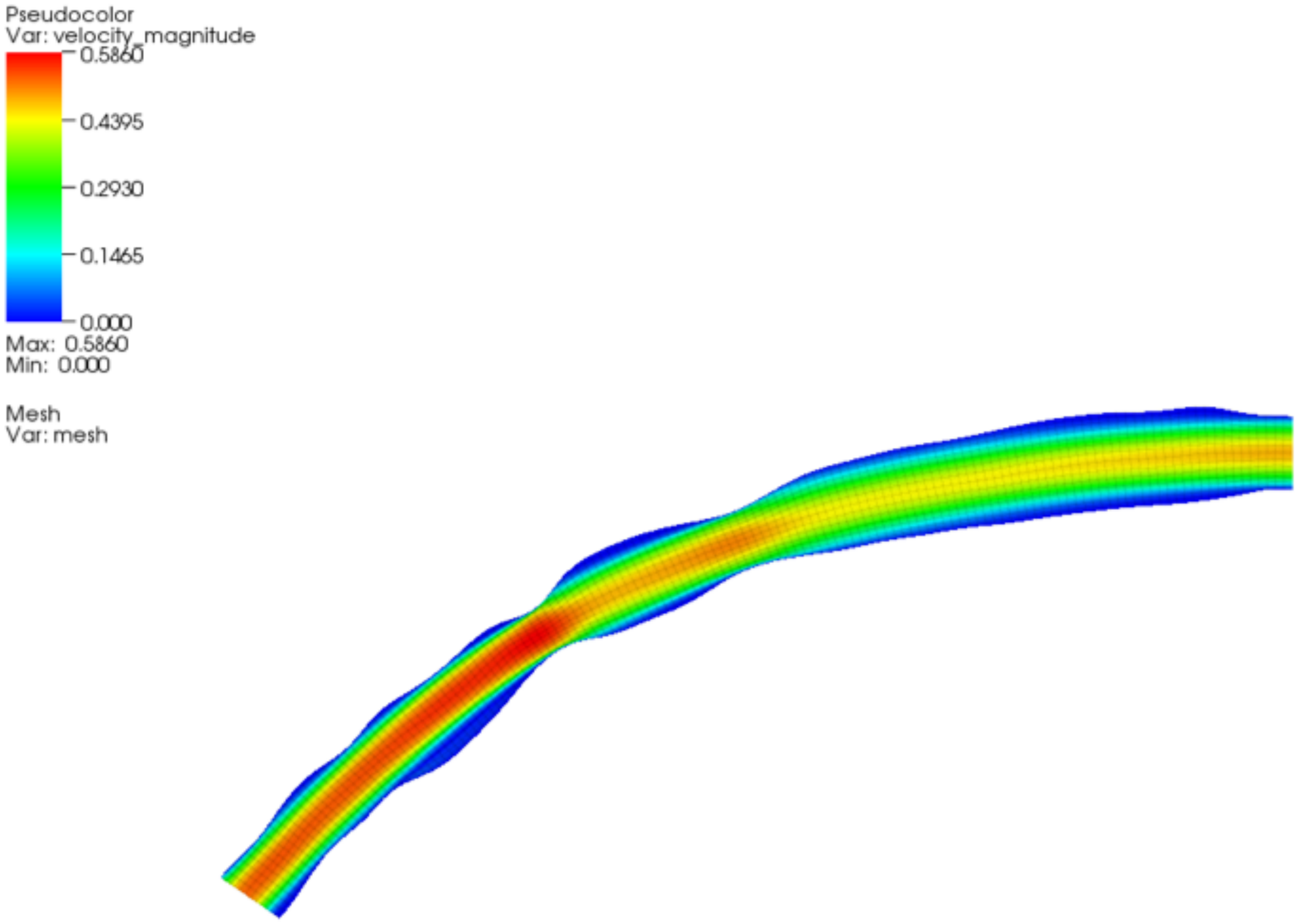}}
 \subfigure[]{\includegraphics[scale=0.3]{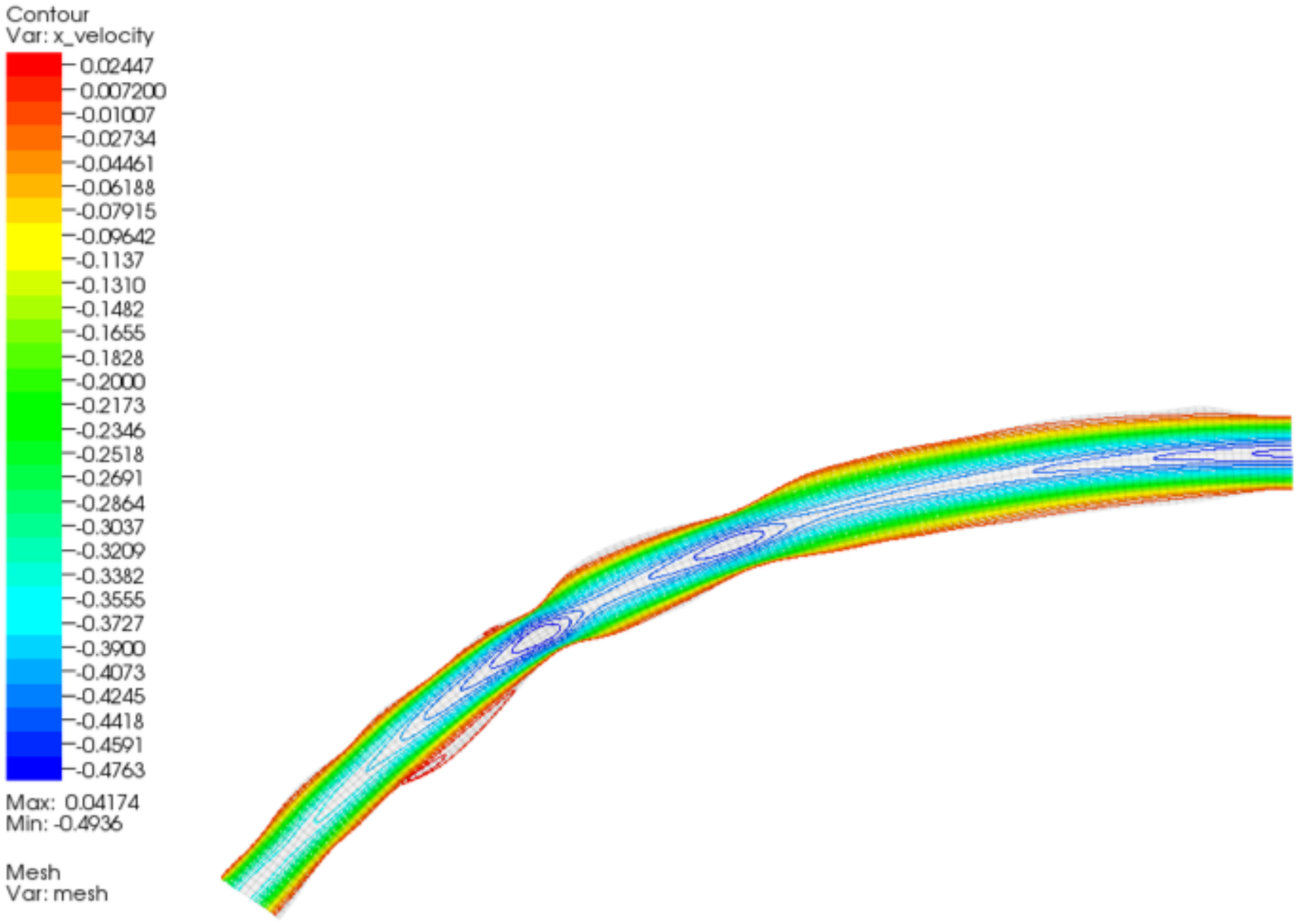}}
 \caption{(a) X velocity and (b) velocity magnitude contours - non-Newtonian simulation with Quemada fitted to the specific viscometry data at $t=1~\mbox{sec} $~ of subject 27 at three month.}
 \label{category5}
\end{figure}

\begin{figure}
\centering 
 \subfigure[]{\includegraphics[height=0.18\textheight]{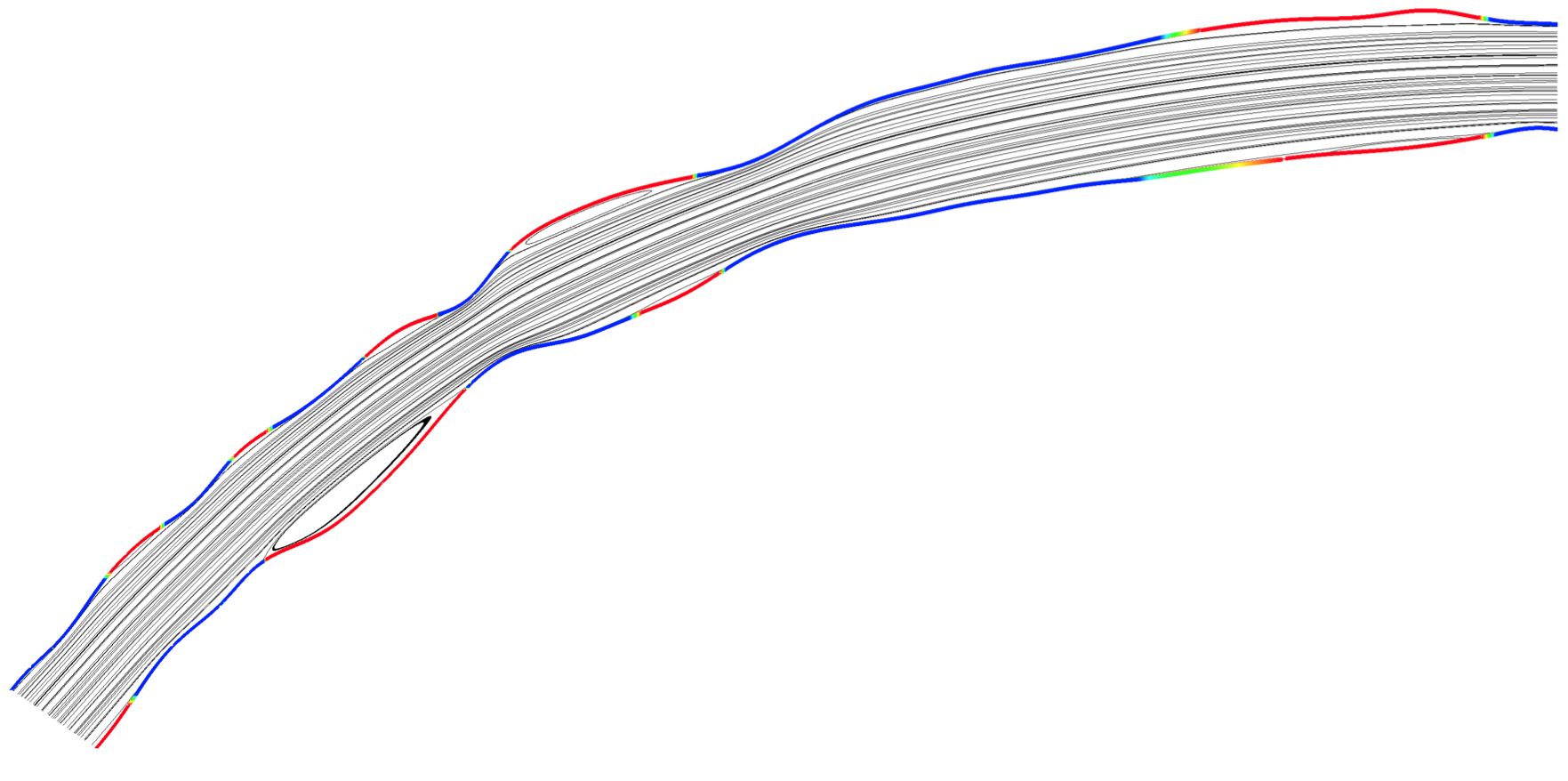}}
 \subfigure[]{\includegraphics[height=0.18\textheight]{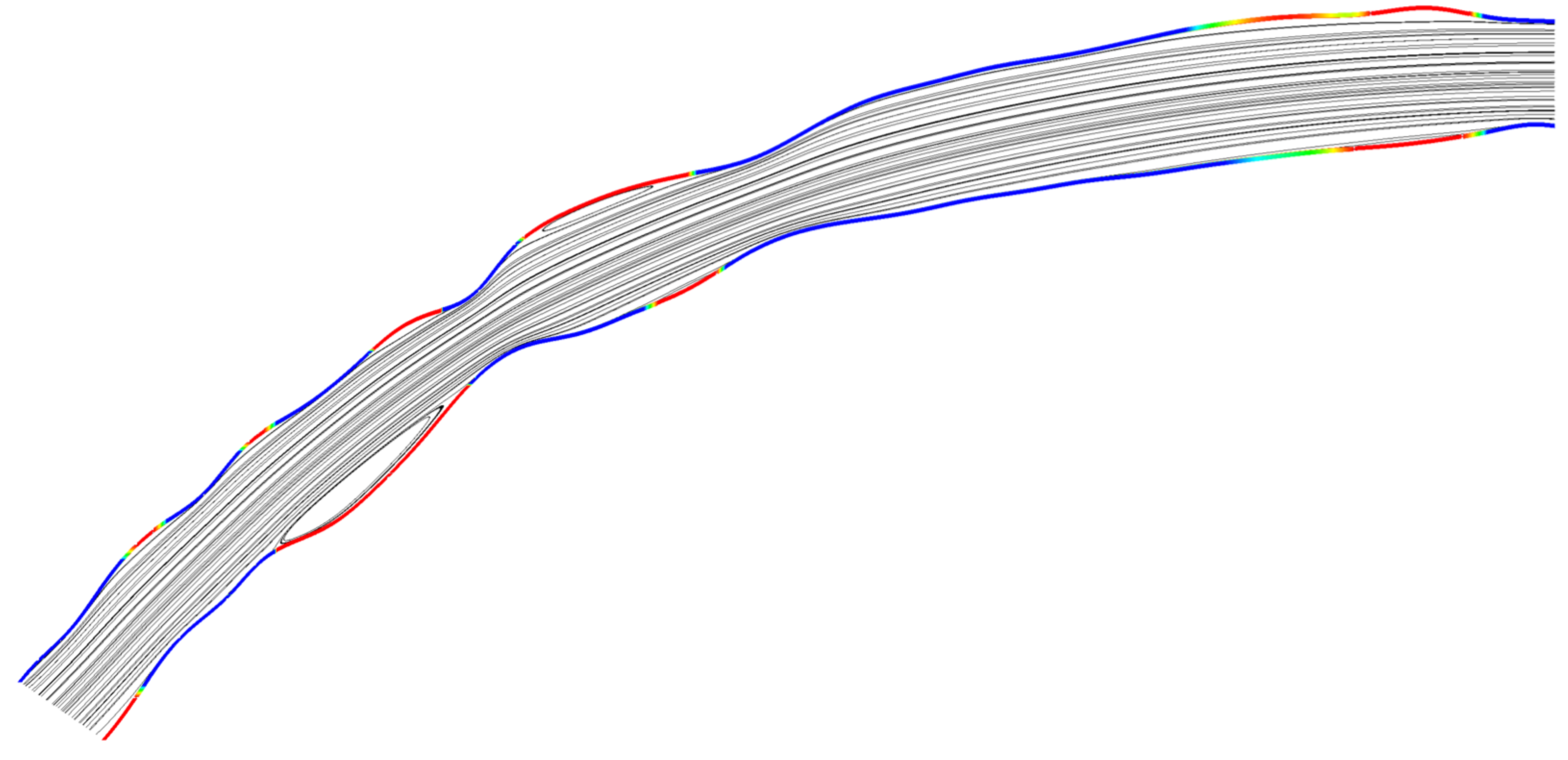}}
  \subfigure[]{\includegraphics[height=0.19\textheight]{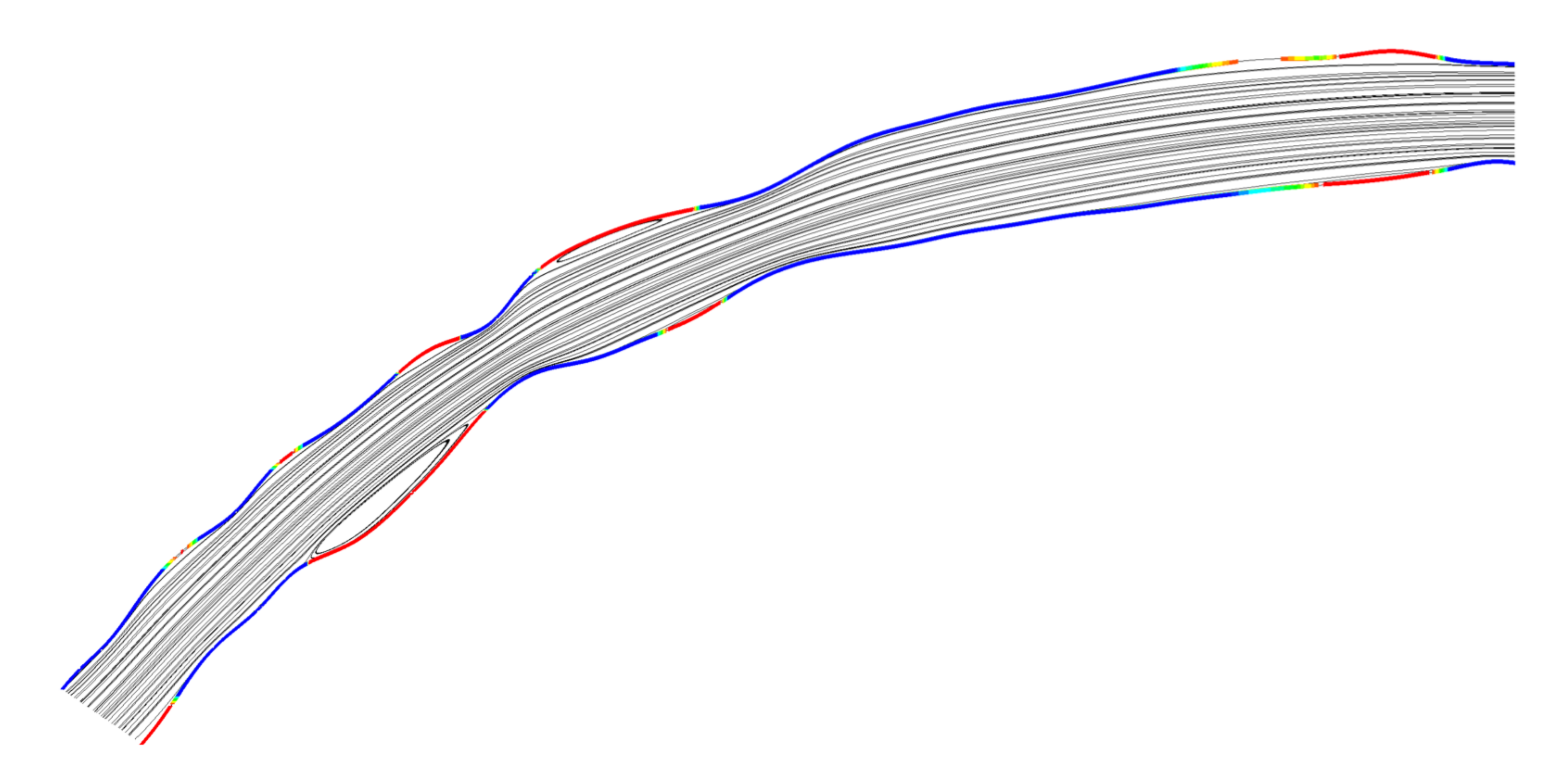}}
 \caption{Distribution of WSS on vessel walls and streamfunction plots for the (a) Newtonian, (b) Quemada, and (c) Casson patient-specific simulations for patient number 27 at 3 month. Thick-red lines are indicative of low-WSS regions where NH could be triggered. Notice the difference in red-zones between Newtonian and non-Newtonian simulations. Green color represents the gradient of WSS values to physiologically acceptable values indicated by a blue line. }
 \label{wss-distribution}
\end{figure}

\begin{figure}
\centering 
 \subfigure[]{\includegraphics[height=0.22\textheight]{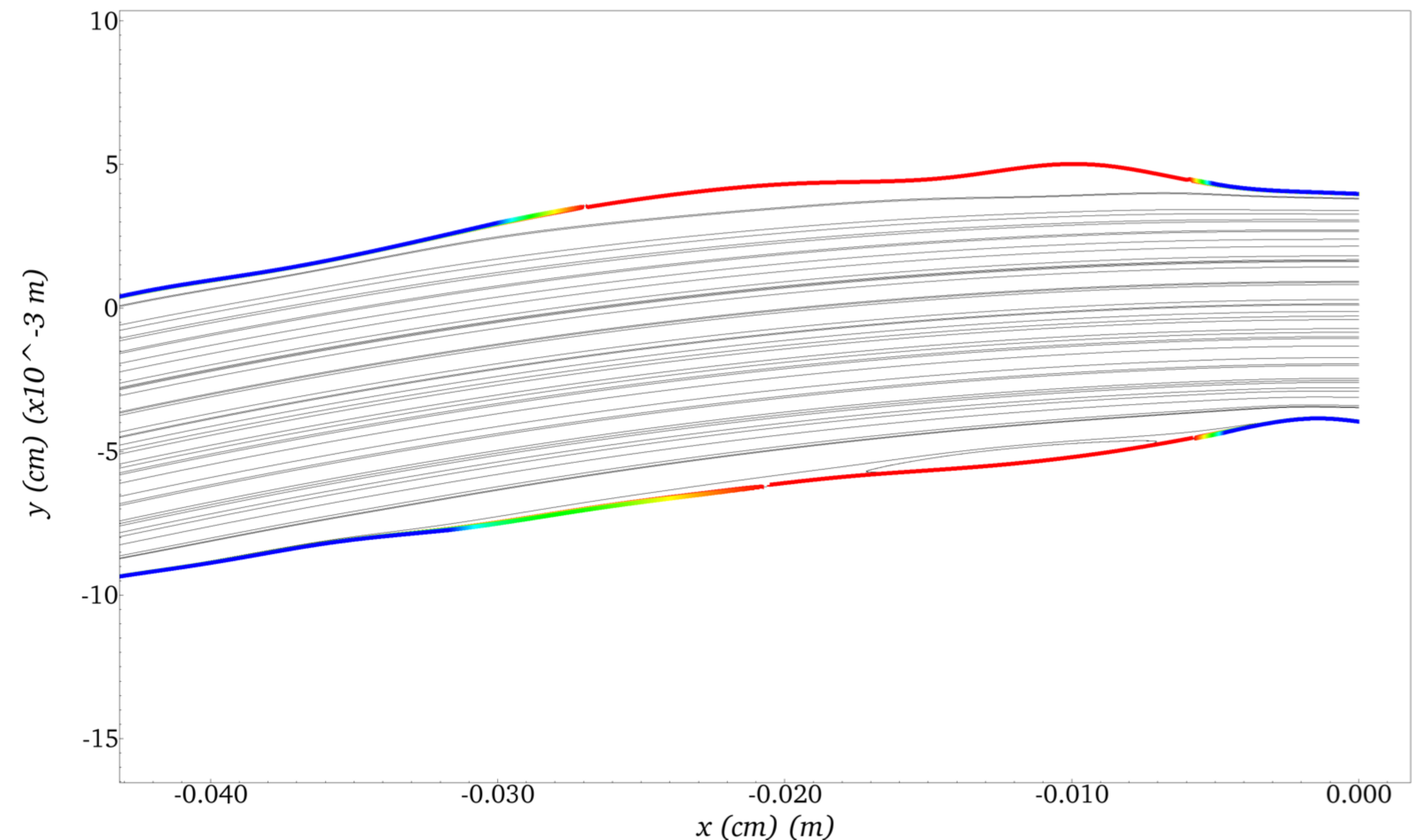}}
 \subfigure[]{\includegraphics[height=0.22\textheight]{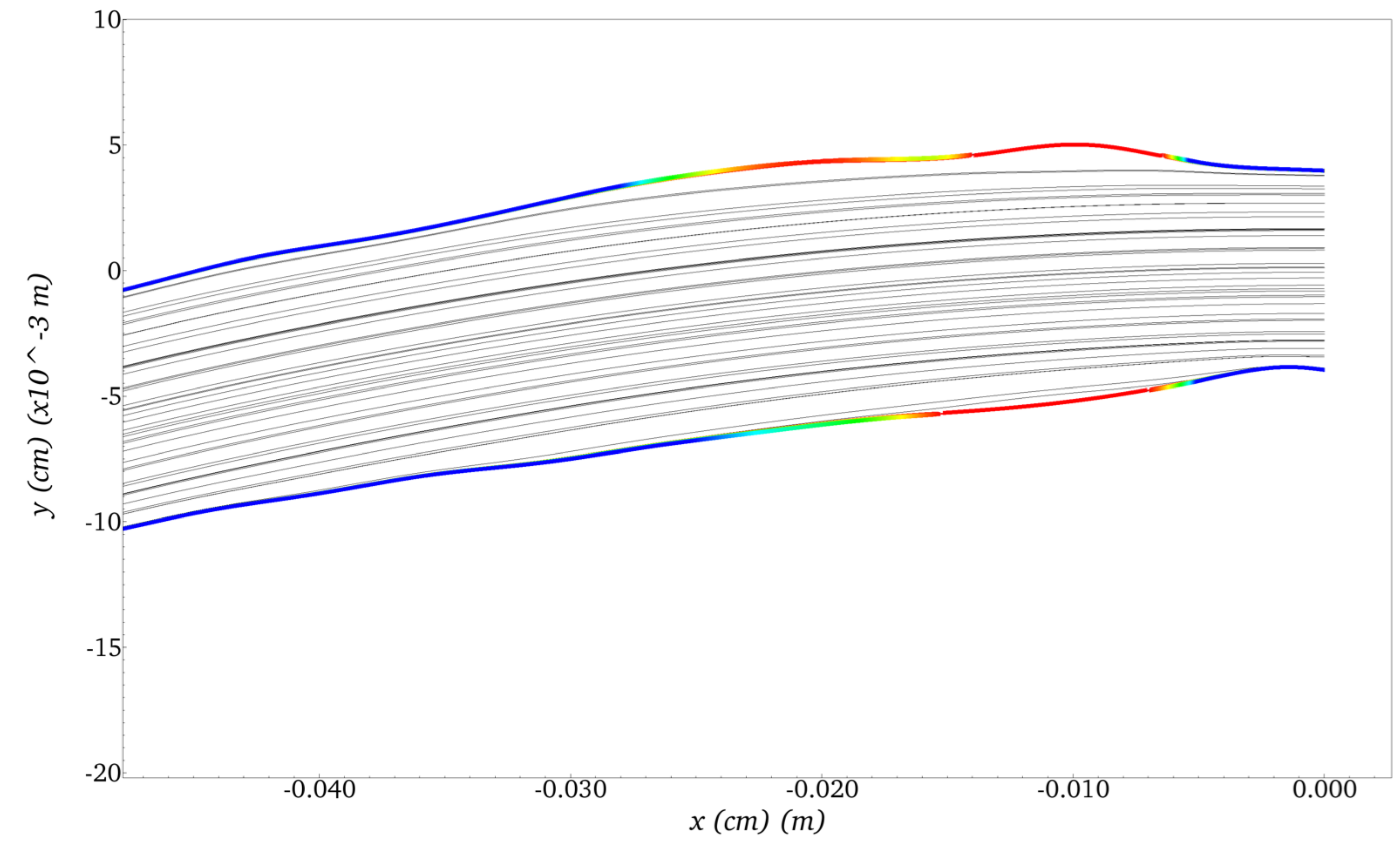}}
  \subfigure[]{\includegraphics[height=0.22\textheight]{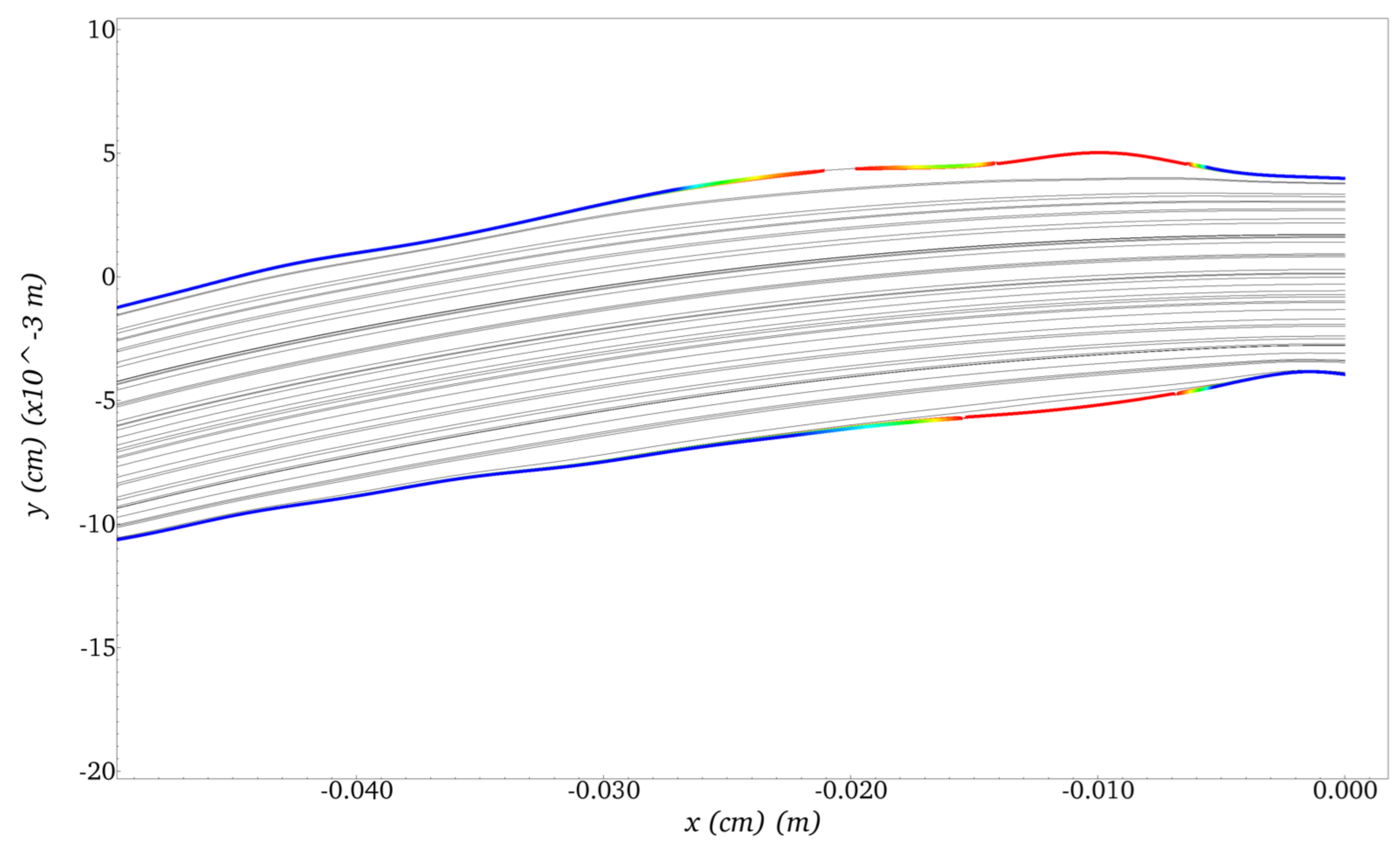}}
 \caption{Distribution of WSS in the inlet region (a) Newtonian, (b) Quemada, and (c) Casson patient-specific simulations. Thick-red lines are indicative of low-WSS regions where NH could be triggered. Notice the difference in red-zones between Newtonian and non-Newtonian simulations.}
 \label{wss-distribution-inlet}
\end{figure}

\begin{figure}
\centering 
 \subfigure[]{\includegraphics[height=0.25\textheight]{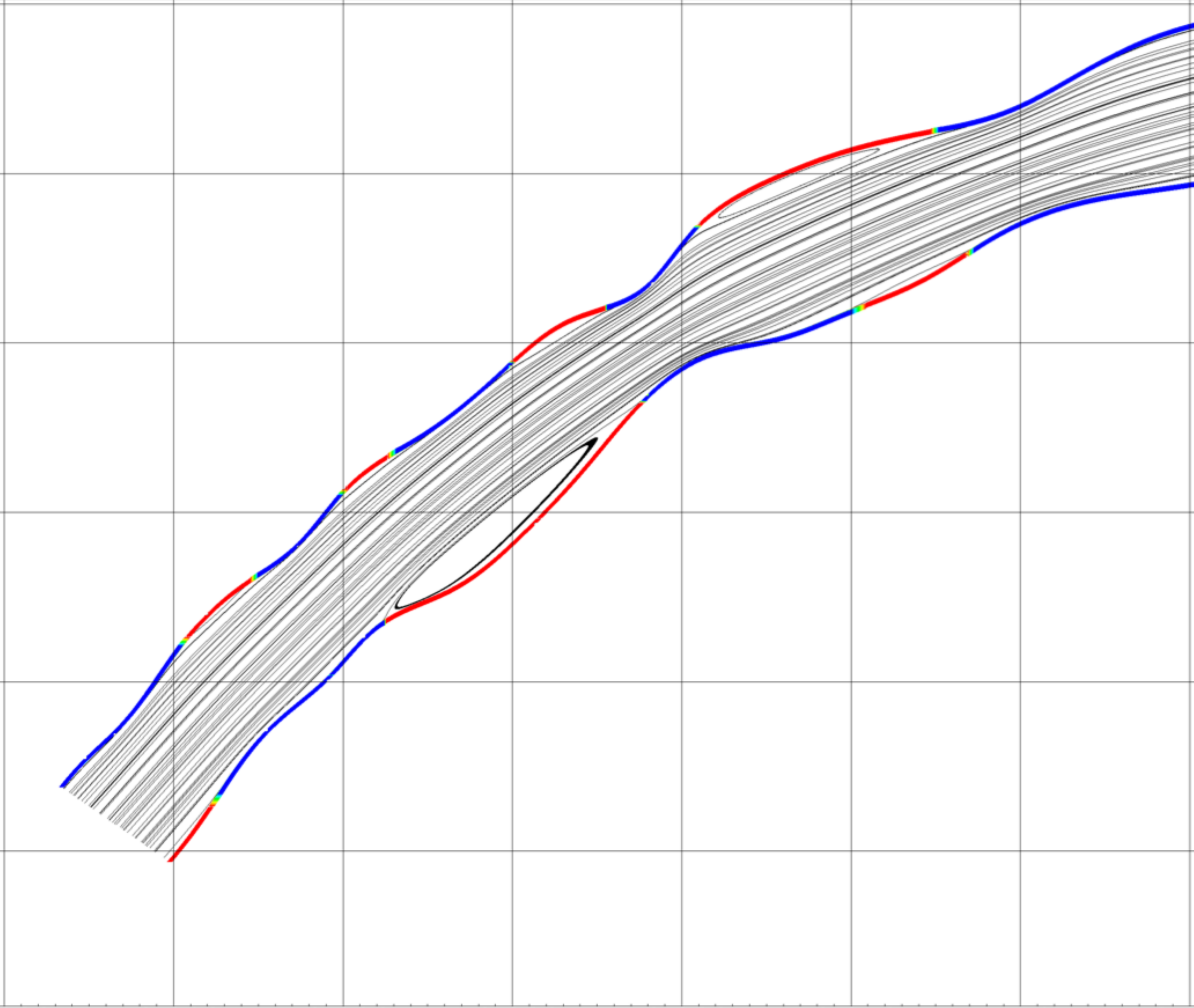}}
 \subfigure[]{\includegraphics[height=0.23\textheight]{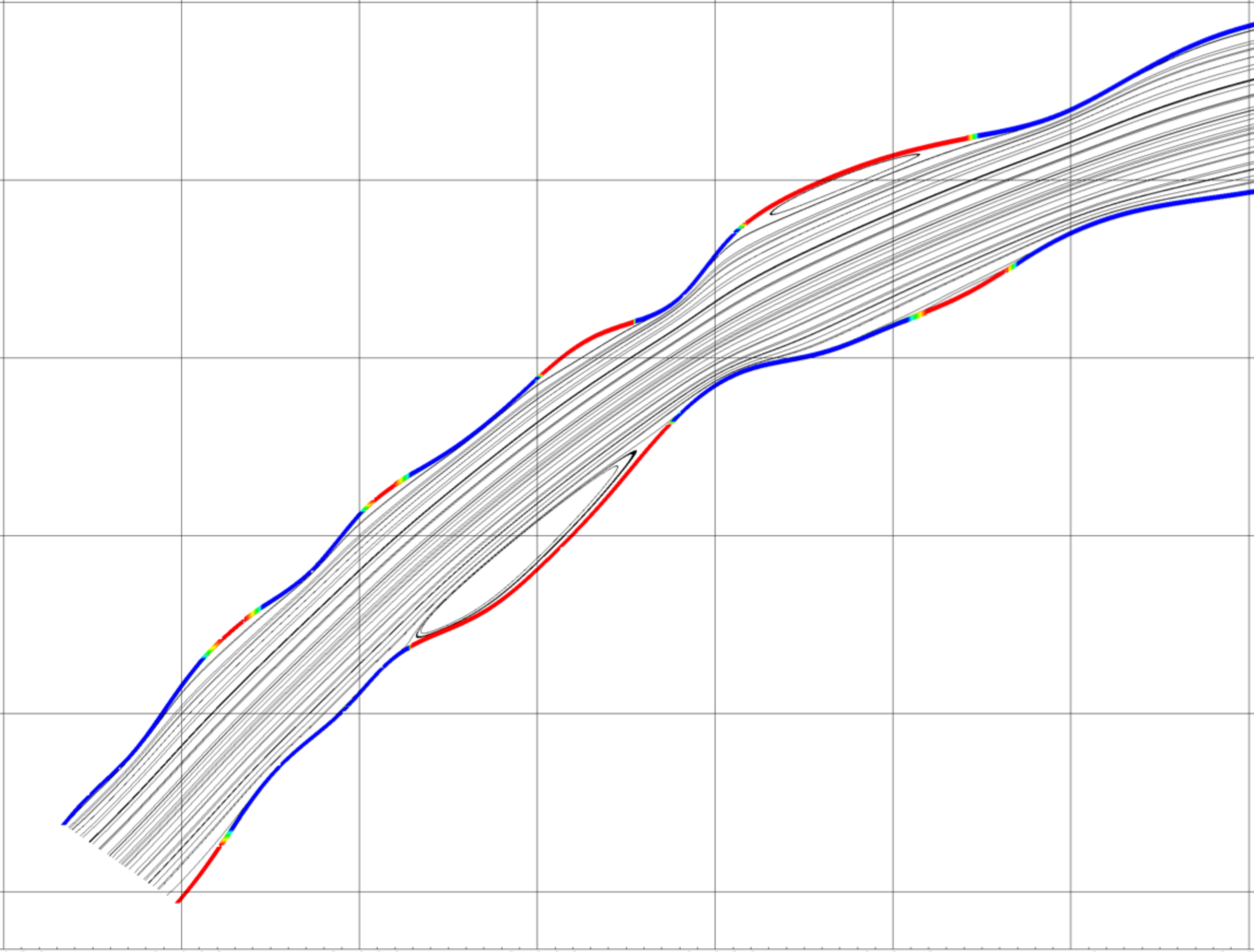}}
  \subfigure[]{\includegraphics[height=0.232\textheight]{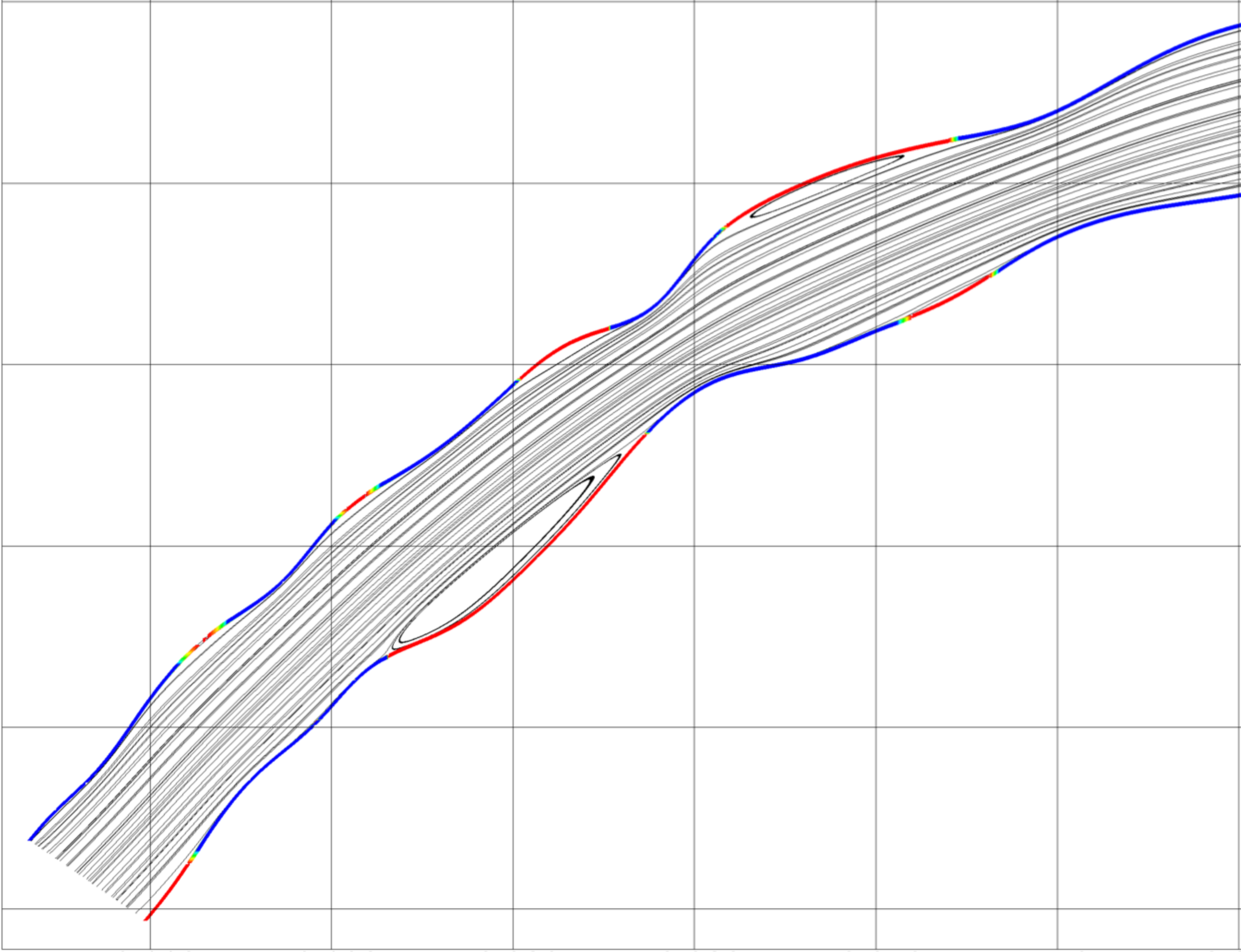}}
 \caption{Distribution of WSS in the mid-region (a) Newtonian and (b) non-Newtonian simulation. Thick-red lines are indicative of low WSS regions where NH could be triggered.}
 \label{wss-distribution-kamar}
\end{figure}

As mentioned before, a comparison of CFD results between categories two and six will make the differences between Newtonian and non-Newtonian effects on the WSS clear. As for the minimum WSS occurring in these two categories, Newtonian simulation predicts the lowest WSS to be -1.16525 Pascal while the non-Newtonian patient specific Quemada yields -0.964495 Pascal. This is a 17.7\% difference between the minimum WSS predicted by a Newtonian and a non-Newtonian simulations. It is documented in the literature that threshold value for WSS to initiate NH is 0.076 Pascal \citep{VanTricht}. In figures~\ref{wss-distribution}, \ref{wss-distribution-inlet}, and \ref{wss-distribution-kamar} these locations are marked with thick-red lines. It can be seen that larger areas in the vessel in a Newtonian flow are endangered by NH (have a WSS value lower than 0.076 $\left[\mbox{Pa}\right]$).

Our simulations revealed that, 46.045\% of the vein and 45.839\% of the arch was affected by low WSS regions in the Newtonian flow case. These numbers drop to 40.316\% and 42.190\% for Quemada and 37\% and  39\% for Casson simulations with the exact same numerical grid, respectively.  Therefore, it was concluded that the implementation of a Newtonian constant reference viscosity is not adequate to predict the risk of NH in ESRD patients (consult figures~\ref{wss-distribution},~\ref{wss-distribution-inlet}, and \ref{wss-distribution-kamar}).  

It has to be pointed out that a high value of WSS also can be harmful to the endothelial cells lining the blood vessels. However, in the cases studied in this research, the maximum values obtained were significantly lower than the maximum threatening threshold (roughly 30 Pascal). As mentioned earlier, no significant difference was observed between Quemada and Casson models. Both models are very similar when implemented in CFD simulation of ESRD patients' vascular systems. We suspect that the effects of the geometry on the hemodynamics are far more important than the non-Newtonian effects. Same was suggested by \citep{chandar}.

Although the implementation of non-Newtonian models in the CFD practice seemed to require more effort from a pragmatic stand point, they manifested computationally less expensive. The dissipative non-Newtonian effects did modify the flow in such way that it took fewer iterations per time step to converge, as compared to Newtonian simulation. Therefore, non-Newtonian simulations took comparable CPU time, despite the added computational cost on the non-Newtonian constitutive relations calculations.  Marrero et. al. \citep{marrero} reported a similar finding in their article.

\section{SUMMARY AND CONCLUSIONS}

It is concluded that the volume concentration, known as hematocrit, has to be considered as an independent variable that affects the viscosity directly. A thorough search in the literature revealed that there are only a couple  non-Newtonian models presented that encompasses both shear rate and hematocrit as independent variables. Therefore, Quemada and Casson models were chosen to implement in this study. HRP for these functions were generated using a least-squares regression technique to the empirical rheologic data obtained from a viscometry test. These models then were used in a multiple-stage CFD simulation. Based on these arguments, the authors find Quemada the most appropriate two-variable model to be used in CFD simulation of ESRD patients vascular systems.

In the first stage, the Quemada model was used four different ways to consider all the possibilities that could potentially yield differences as compared to a Newtonian simulation. The same was done with the Casson model. The simulations were performed on patient-specific geometries and the Reynolds number considered was 500 for Newtonian cases with the same inlet velocity profile for non-Newtonian simulations.

It is expected that the effects of viscous forces are more pronounced in non-Newtonian flows. Hence, thicker viscous sublayers and flatter velocity profiles are expected in these flows as well as less frequent occurrence of secondary flow zones, similar to the reported results of Cherry and Eaton \citep{cherry}. The largest differences were observed in minimum pressure (almost 43\%) and in X velocity (28\%). For the WSS, the results indicate that the Newtonian simulation may over-predict the vulnerability to neointimal hyperplasia and onset of stenotic sites. The difference in WSS values were up to 17\% between Newtonian and non-Newtonian cases considered here, which is a considerable amount since vessel walls and endothelial cells physiologically are very sensitive to this parameter. The combination of non-physiological flows (i.e. very high flow rates) with the curvature of the vessels increases the likelihood of low-WSS regions and susceptibility to NH and eventually CAS. 

 It is also noted that the influences of the geometry may play a more important role in hemodynamics than the non-Newtonian effects. This is the first study to evaluate the influence of non-Newtonian effects in patient-specific geometries using patient-specific viscosity data for a large cohort. The primary shortcoming of the investigation is the use of two-dimensional geometries which cannot capture the secondary flows common in three-dimensional curved tubes and vessels.

\vspace{0.1in}

\bibliographystyle{model2-names}
\bibliography{paperBiblio}

\end{document}